\documentclass[aip,apl,reprint]{revtex4-1}

\pdfoutput=1
\usepackage{graphicx}
\usepackage{amssymb}
\usepackage{amsmath}
\usepackage[utf8]{inputenc}
\usepackage[T1]{fontenc}

\newcommand{\Lk}[1] {
L_\mathrm{k#1} }
\newcommand{\Istar}[1] {
I^{*#1} }
\newcommand{\Is}{I_\mathrm{s}}
\newcommand{\Ic}{I_\mathrm{c}}
\newcommand{\It}{I_\mathrm{T}}
\newcommand{\Qt}{Q_\mathrm{t}}
\newcommand{\Qe}{Q_\mathrm{e}}
\newcommand{\Qi}{Q_\mathrm{i}}
\newcommand{\vin}{V_\mathrm{in}}
\newcommand{\Ltot}{L_\mathrm{tot}}
\newcommand{\Lg}{L_\mathrm{g}}
\newcommand{\alphak}{\alpha_\mathrm{k}}
\newcommand{\Prf}{P_\mathrm{rf}}

\newcommand{\Tc}{T_\mathrm{c}}
\newcommand{\Jc}{J_\mathrm{c}}
\newcommand{\fr}{f_\mathrm{r}}
\newcommand{\Tmin}{T_\mathrm{min}}

\begin{document}


\title{Magnetic field sensing with the kinetic inductance of a high-$\Tc$ superconductor} 

\author{V. Vesterinen}
\email[]{visa.vesterinen@vtt.fi}
\affiliation{VTT Technical Research Centre of Finland Ltd, QTF Center of Excellence, P.O. Box 1000, FI-02044 VTT, Finland}

\author{S. Ruffieux}
\author{A. Kalabukhov}
\affiliation{Department of Microtechnology and Nanoscience—MC2, Chalmers University of Technology, SE-41296 Gothenburg, Sweden}

\author{H. Sipola}
\author{M. Kiviranta}
\affiliation{VTT Technical Research Centre of Finland Ltd, QTF Center of Excellence, P.O. Box 1000, FI-02044 VTT, Finland}

\author{D. Winkler}
\affiliation{Department of Microtechnology and Nanoscience—MC2, Chalmers University of Technology, SE-41296 Gothenburg, Sweden}

\author{J.~F. Schneiderman}
\affiliation{MedTech West and the Institute of Neuroscience and Physiology, University of Gothenburg, SE-40530 Gothenburg, Sweden}

\author{J. Hassel}
\affiliation{VTT Technical Research Centre of Finland Ltd, QTF Center of Excellence, P.O. Box 1000, FI-02044 VTT, Finland}

\date{\today}

\begin{abstract}
We carry out an experimental feasibility study of a magnetic field sensor based on the kinetic inductance of the high-$\Tc$ superconductor yttrium barium copper oxide. We pattern thin superconducting films into radio-frequency resonators that feature a magnetic field pick-up loop. At 77~K and for film thicknesses down to 75~nm, we observe the persistence of screening currents that modulate the loop kinetic inductance. According to the experimental results the device concept appears attractive for sensing applications in ambient magnetic field environments. We report on a device with a magnetic field sensitivity of $4$~pT$/\sqrt{\mathrm{Hz}}$, an instantaneous dynamic range of $11~\mu$T, and operability in magnetic fields up to $28~\mu$T. 
\end{abstract}

\pacs{}

\maketitle 

The kinetic inductance of superconductors has found many applications in fields as diverse as bolometry~\cite{timofeev_optical_2017,lindeman_arrays_2014}, parametric amplification~\cite{ho_eom_wideband_2012,ranzani_kinetic_2018}, current detectors~\cite{wang_kinetic_2018}, and sensing of electromagnetic radiation~\cite{rantwijk_multiplexed_2016,sato_evaluation_2018}, to name but a few. Each device harnesses a certain type of a non-linearity of the kinetic inductance $\Lk{}$, such as that induced by temperature, electric current, or non-equilibrium quasiparticles. In sensor applications, radio-frequency (rf) techniques are often employed in observation of the variations of $\Lk{}$: a high sensitivity follows from the intrinsically low dissipation of the superconductors, manifesting itself as a high quality factor of resonator circuits, for example.

The general advantages common to all $\Lk{}$ sensors have motivated the development of kinetic inductance magnetometers (KIMs), devices that combine the $\Lk{}$ current non-linearity with magnetic flux quantization~\cite{luomahaara_kinetic_2014,asfaw_skiffs:_2018}. In comparison to the most established category of sensitive magnetometers, i.e., superconducting quantum interference devices (SQUIDs), KIMs have certain benefits. KIM fabrication involves only a single-layer process that avoids Josephson junctions which are the central components of SQUIDs. Furthermore, KIMs typically have a higher dynamic range, and they enable operation in demanding ambient magnetic field conditions. KIMs also unlock the ability to use frequency multiplexing for the readout of large sensor arrays where each magnetometer has a dedicated eigenfrequency~\cite{rantwijk_multiplexed_2016,sipola_multiplexed_2018}.

In this Letter, we demonstrate KIMs fabricated from yttrium barium copper oxide (YBCO). YBCO is a high-$\Tc$ superconductor that enables KIM operability at elevated temperatures $T$, allowing for cooling with liquid nitrogen. Non-linear $\Lk{}$ of YBCO has previously been evaluated for bolometric~\cite{lindeman_arrays_2014} (direct~\cite{sato_evaluation_2018}) detection of infrared (optical) radiation. A further benefit of the material is its high tolerance against background magnetic fields, which has recently culminated in a YBCO rf resonator with a quality factor of about $10^4$ at $T<55~$K and at a magnetic flux density of $7~$T applied parallel to the superconducting film~\cite{ghirri_yba2cu3o7_2015}. From a sensitivity viewpoint, an important benchmark for our KIM are state-of-the-art YBCO SQUID magnetometers~\cite{faley_high-_2017} that have a sensitivity better than $50~\mathrm{fT}/\sqrt{\mathrm{Hz}}$. However, these SQUIDs suffer from a complicated fabrication process that makes mass production difficult, and in order to extend the dynamic range beyond a few nT, they need to be operated in a flux-locked loop requiring at least four wires to each cold sensor.

We review the KIM operating principle, starting from the $\Lk{}$ current non-linearity~\cite{zmuidzinas_superconducting_2012,vissers_frequency-tunable_2015},
\begin{equation}
\Lk{} = \Lk{0}[ 1 + \left( \Is/\Istar{}\right)^2 ] .
\label{eq:nonlinear}
\end{equation}
In anticipation of using it for magnetometry, we have plugged in a screening current $\Is$, the flow of which is enforced by magnetic flux quantization. $\Lk{0}$ is the kinetic inductance at $\Is=0$, and $\Istar{}$ is a normalizing current on the order of the critical current $\Ic$. Assuming that $\Lk{}$ is a property of a superconducting loop with an area $A$, we formulate the flux quantization as
\begin{equation}
(\Lg+\Lk{})\Is - B_0A= m\Phi_0 ,
\label{eq:quantize}
\end{equation}
where $\Lg$ is loop geometric inductance, $B_0$ the spatial average of the magnetic flux density threading the loop, and $m\Phi_0$ an integer times the magnetic flux quantum.  Sensitive magnetometry calls for a decent kinetic inductance fraction $\alphak = \Lk{}/(\Lg+\Lk{})$, and an effective method of observing the $B_0$-induced inductance variations. To establish an rf readout, two opposite edges of the loop are connected with a capacitor that leaves $\Is$ unperturbed, but creates an rf eigenmode together with the loop inductance. Then, the inductance variation translates into a changing resonance frequency, a quantity which is probed by coupling the resonator weakly into a 50-$\Omega$ readout feedline. Two KIMs of this kind have recently been reported: the materials of choice have been NbN~\cite{luomahaara_kinetic_2014} and NbTiN~\cite{asfaw_skiffs:_2018}, both of which are low-$\Tc$ superconductors whose disordered nature provides a magnetic penetration depth~\cite{vissers_low_2010} $\lambda$ exceeding several hundreds of nm. For films with a thickness $d\ll \lambda$, the kinetic surface inductance equals $\mu_0 \lambda^2/d$, with $\mu_0$ the vacuum permeability.

In the design of our high-$\Tc$ KIM [Fig.~1(a)], we use as a guideline the theoretical responsivity on resonance~\cite{luomahaara_kinetic_2014}
\begin{equation}
\left| \frac{\partial V}{\partial B_0} \right| = \frac{\Qt^2 \vin \Is A}{4\Qe \Ltot \Istar{2} [1/\alphak +3(\Is/\Istar{})^2]}
\label{eq:resp}
\end{equation}
that describes how the magnetic flux sensitivity of the resonator voltage $V$ is related to the electrical and geometric device parameters. The readout rf power $\Prf$ is expressed through an excitation voltage amplitude $\vin$. The total inductance $\Ltot$ equals one quarter of the loop inductance plus the contribution of the parasitic trace connecting the two halves of the loop. $\Qt$ ($\Qe$) denotes the loaded (external) quality factor. We choose a maximal loop area allowed by fabrication technology $A=(8~\mathrm{mm})^2$. We anticipate that reaching a significant $\alphak$ is the main bottleneck: The reported values~\cite{ghigo_microwave_2004} of the YBCO $\lambda$ in the low-$T$ limit are only $150-200$~nm, and few devices have previously featured long superconducting traces with a small cross-section~\cite{hattori_5-/spl_1998}. We select a conservative value of $w=10~\mu$m, and compare devices with a variable $d\leq 225~$nm. The shunt capacitor $C \simeq 16~$pF, which determines the unloaded angular resonance frequency through the relation $(\Ltot C)^{-1/2}$, is formed from interdigitated fingers of width $10~\mu$m and gap $5~\mu$m. We target a loaded resonance frequency $\fr$ of about $250$~MHz.

\begin{figure}
\includegraphics{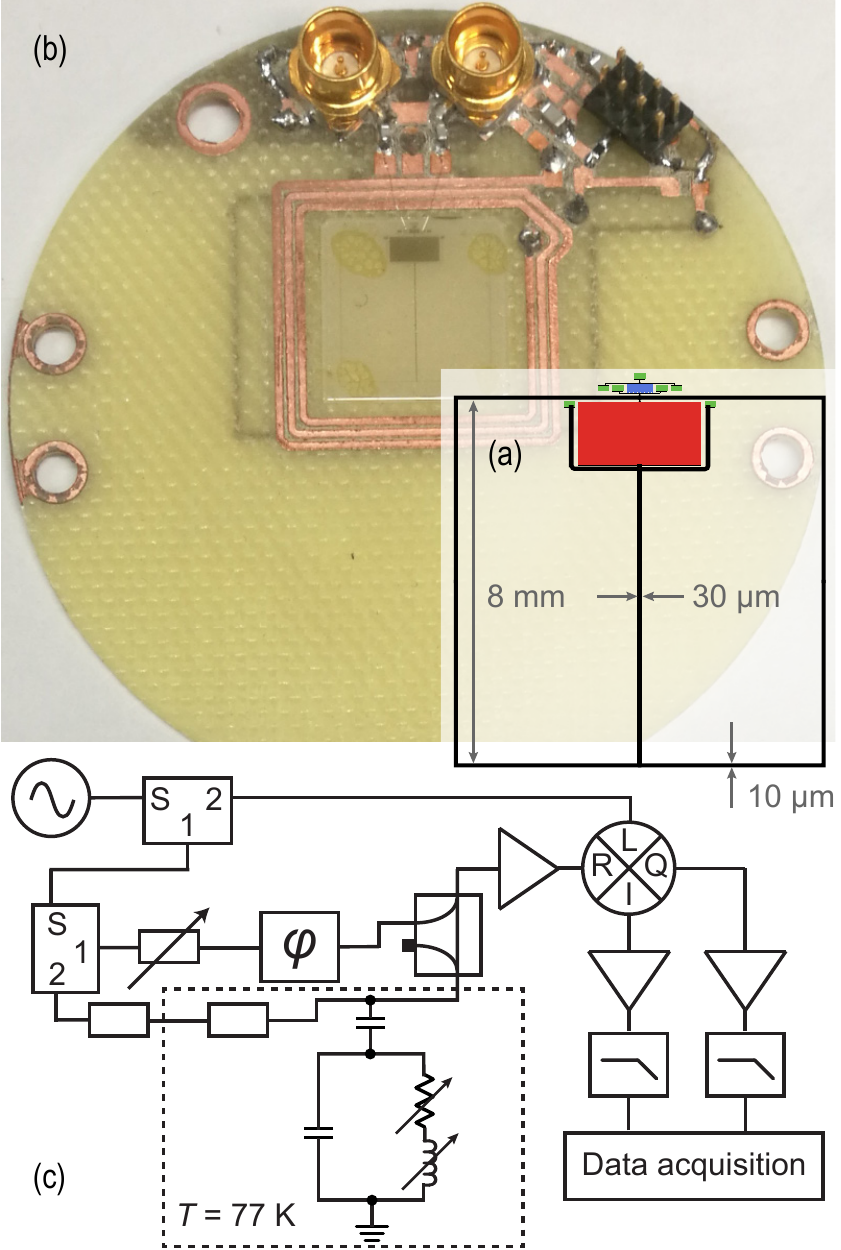}%
\caption{\label{fig:Fig1}Magnetometer and its readout scheme. (a) The superconductor mask layout of Sample B with an on-chip shunt capacitor, a coupling capacitor, and bondpads highlighted in red, blue and green, respectively. \emph{The trace widths are not to scale.} (b) A photograph of a magnetometer chip (Sample B) on a sample holder PCB. The 44-mm-diameter PCB hosts a three-turn dc bias coil on its top surface, and a single-turn ac bias coil at the bottom. The calculated mutual inductance between the superconducting loop and the dc (ac) coil is 20~nH (5.3~nH).  (c) A simplified rf readout schematic where three copies of an rf carrier are taken to ensure phase stability. The resonator encodes the magnetic signal into the sidebands of the first copy at low temperature $T$. Before amplification, the (optional) second copy interferometrically cancels the carrier power. The third copy is a reference used in the demodulation of the magnetic signal to dc.}%
\end{figure}

Sample A was fabricated on a $10\times10\text{ mm}^2$ r-cut sapphire substrate with a yttria-stabilized zirconia (YSZ) and a CeO$_2$ buffer layer to support the epitaxial growth of a $d=225$~nm YBCO film. As the deposition process of thin YBCO films is optimized for MgO, Sample B (Sample C) with a $d=75$~nm ($d=50$~nm) YBCO film was fabricated on a 110 MgO substrate without buffer layers instead. Low-loss microwave-resonant YBCO structures have previously been reported on both substrate materials~\cite{ghirri_yba2cu3o7_2015,sato_evaluation_2018}. The YBCO films were deposited with pulsed laser deposition (PLD) and the devices were then patterned with optical lithography using a laser writer and argon ion beam etching. For Samples A and C a simple S1318 photoresist mask was used, while for Sample B a hard carbon mask was chosen to avoid degradation of the thin superconducting film by the photoresist. The carbon mask was deposited with PLD and patterned by oxygen plasma etching through a Cr mask.  The ion beam etching process was monitored with secondary ion mass spectrometry for endpoint detection. We attach the KIMs onto a printed circuit board (PCB) that has copper patterns for rf wiring and magnetic bias coils~[Fig.~1(b)]. 

The first KIM characterization is the measurement of the resonance lineshape and its sensitivity to the magnetic field. As $\Lk{}$ makes the resonator a sensitive thermometer~\cite{lindeman_arrays_2014}, we resort to immersion cooling in liquid nitrogen at $T=77$~K. We use a high-permeability magnetic shield that not only protects the sample from magnetic field noise, but also prevents trapping of flux vortices during the time when the sample crosses $\Tc$. The core elements of the readout electronics are a low-noise rf preamplifier~\cite{noauthor_see_nodate} followed by a demodulation circuit (IQ mixer), and analog-to-digital converters [Fig.~1(c)]. We sweep the frequency of a weak ($\Prf \leq$~-66 dBm) rf tone across the resonance. We simultaneously apply a static $B_0$ as well as a weak ac probe tone at a frequency of 1~kHz. From the averaged in-phase (I) and quadrature (Q) components of the output we extract the complex-valued transmission parameter $S_{21}=2V/\vin$.  In addition, we use ensemble averaging of the modulated output voltage to extract the responsivity $\partial V/\partial B_0$ corresponding to the probe tone.
\begin{figure}
\includegraphics{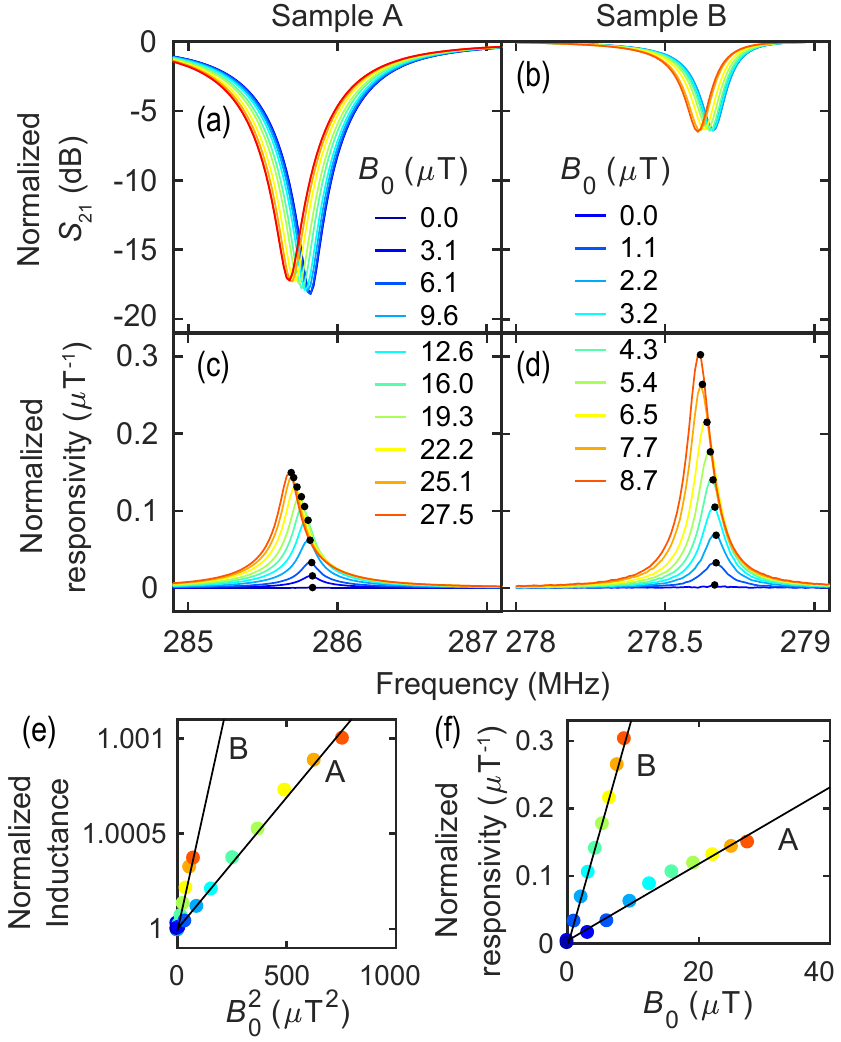}
\caption{\label{fig:Fig2}Characterization as functions of readout frequency and static magnetic field $B_0$. The measurements of Sample A [(a),(c)] and Sample B [(b),(d)] show a qualitatively similar response to a variable $B_0$ (colors) induced by the dc bias coil. (a),(b) The dip in the transmission S-parameter magnitude  $20\log_{10}|2V/\vin|$ gives information on the changes in the $\Lk{}$ and dissipation in the superconducting loop. (c),(d) The responsivity  $|\frac{\partial V}{\partial B_0}/V|$ is extracted from the simultaneous measurement of a 1-kHz magnetic probe tone produced by the ac bias coil. The Sample A (B) probe magnitude is 140~nT (30~nT). The responsivity maxima are indicated with dots. (e) The shift in the resonance frequency is converted into a normalized change in inductance, and presented as a function of $B_0^2$. (f) Maximal responsivity as a function of $B_0$. The solid lines in (e) and (f) are linear fits.}
\end{figure}

From the best fits~\cite{megrant_planar_2012} to $S_{21}$ we extract $\fr$, $\Qe$, and the internal quality factor $\Qi$. Applying $B_0$ shifts $\fr$ downwards in Samples A and B~[Fig.~2(a,b)], but not in Sample C which has the thinnest film. We suspect that a residual dc resistance prohibits the proper flux quantization. However, also Sample C reacts to an ac magnetic excitation, and at a lower $T\approx 60~$K we observe the proper dc response as well. We convert the frequency shifts of Samples A and B into an equivalent change in $\Ltot$ and observe a quadratic dependence on $B_0$ [Fig.~2(e)], which is in line with Eqs.~(\ref{eq:nonlinear}-\ref{eq:quantize}). Unlike in low-$\Tc$ KIMs~\cite{luomahaara_kinetic_2014,asfaw_skiffs:_2018}, we do not observe reseting of the sample to $\Is=0$ [i.e., into a finite $m$ in Eq.~(\ref{eq:quantize})] upon crossing a threshold $B_0$ corresponding to $\Is=\Ic$. Instead, the resonance of Sample A (B) stays put at $B_0\geq 28~\mu\mathrm{T}$ ($B_0 \geq 9~\mu\mathrm{T}$). We attribute this to flux trapping that most likely occurs at the sample corners where the inhomogeneous bias field of the square coil is the strongest (about $1.5 B_0$). Ref.~\onlinecite{sun_initial-vortex-entry-related_1994} proposes a flux-trapping condition of the form $\Is \geq \It \propto (\Jc d)^{3/4} w^{1/2}$ where $\Jc$ is the critical current density.

Regarding the quality factors, we note that Sample A is overcoupled with $\Qe = 350$ much smaller than $\Qi \leq 2100$. Sample B is close to being critically coupled ($\Qe= 3500$, $\Qi \leq 3700$). The observed $\Qi$ variations are on the order of $10 \%$. Low internal dissipation is key to achieving high device sensitivity. Thus, we discuss the possible mechanisms affecting $\Qi$. Firstly, the resistive part of the superconductor rf surface impedance generates loss that grows with increasing $T$, $\Is$, and surface roughness~\cite{zaitsev_microwave_2002,wang_growth_2003}. The dielectric losses of the substrates should not play a role: both sapphire and MgO have low relative permittivity (sapphire: $\epsilon_r=9.3$ and $\epsilon_{rz}=11.3$ anisotropic, MgO: $\epsilon_r=9.6$) and low dielectric loss tangents~\cite{krupka_dielectric_1994,buckley_cryogenic_1994,taber_microwave_1995} (<$4 \cdot 10^{-6}$) at $T=77$~K. A loss mechanism related to the PCB deserves further attention: the presence of the bias coils made of resistive copper. In Ref~\onlinecite{luomahaara_kinetic_2014} as well as for the data presented in Figs.~2-3 for Sample A, bias coil rf decoupling is attemped with series impedances (resistance, inductance) on the order of hundreds of Ohms within the coils. This has allowed for $\Qi=1000-2000$, but we have learned that higher values can be reached with an arrangement where the bias coils are grounded at rf and the readout is mediated by stray coupling between the KIM and the coils~\cite{noauthor_see_nodate}. Samples B and C as well as a subsequent cooldown of Sample A (Fig.~4) have been prepared using this new method. 

The device responsivities are presented in Fig.~2(c,d) as a function of the readout frequency. They are of the normalized form $|V^{-1} \partial V/\partial B_0|$: this is a convenient quantity because both $\partial V/\partial B_0$ and V experience the same gain of the readout electronics. The measured readout frequency dependencies are Lorentzians that peak on resonance. We use these data for the estimation of the sensor dynamic range~\cite{luomahaara_kinetic_2014}, which is approximately $11~\mu$T ($2.8~\mu$T) for Sample A (B) at high responsivity. As expected, the responsivity vanishes at the first-order flux-insensitive points where $\Is=0$. The peaks of the responsivity Lorentzians, shown as a collection in Fig.~2(f), are linearly proportional to $B_0$. If we normalize Eq.~(\ref{eq:resp}) in the limit of low $\alphak \ll 1$,
\begin{equation}
\left|\frac{1}{V} \frac{\partial V}{\partial B_0} \right| \approx \frac{\Qi^2  A}{2(\Qi+\Qe)} \frac{\alphak \Is}{\Ltot \Istar{2} } ,
\label{eq:normresp}
\end{equation}
we obtain a model which is in line with the measured trend since there is an approximately linear mapping from $B_0$ into $\Is$ [consider Eq.~(\ref{eq:nonlinear}) at $m=0$]. The measured linear slope of the responsivity is about six times steeper in Sample B, which is primarily an indication of a higher $\alphak$ and a lower $\Istar{}$ resulting from the thinner film.

To determine the magnetic field sensitivity of Samples A and B, we record $1.0~$s voltage traces and average the squared modulus of their Fourier transform. We do this at $\fr$ as a function of $\Prf$: importantly, a carrier cancellation circuit [Fig.~1(c)] is activated now to avoid the saturation of the electronics. We also compare $B_0$ bias points with a high and a vanishingly small responsivity to the magnetic flux. At low $\Prf$ we observe a white voltage spectrum determined by thermal noise and noise added by the preamplifiers. As we increase the power, a $1/f$-like spectrum emerges and eventually dominates the voltage noise~\cite{noauthor_see_nodate}. The voltage spectra at the high responsivity and at the highest rf power have been converted into the magnetic domain in Fig.~3, yielding a sensitivity of about $4~\mathrm{pT}/\sqrt{\mathrm{Hz}}$ at $10~$kHz for both KIMs. We can rule out direct magnetic field noise because the voltage spectrum is similar at the operating point with vanishing responsivity. The origin of the $1/f$ mechanism is currently not fully understood. The $1/f$ output voltage spectral density scales linearly with $\vin$.
\begin{figure}
\includegraphics{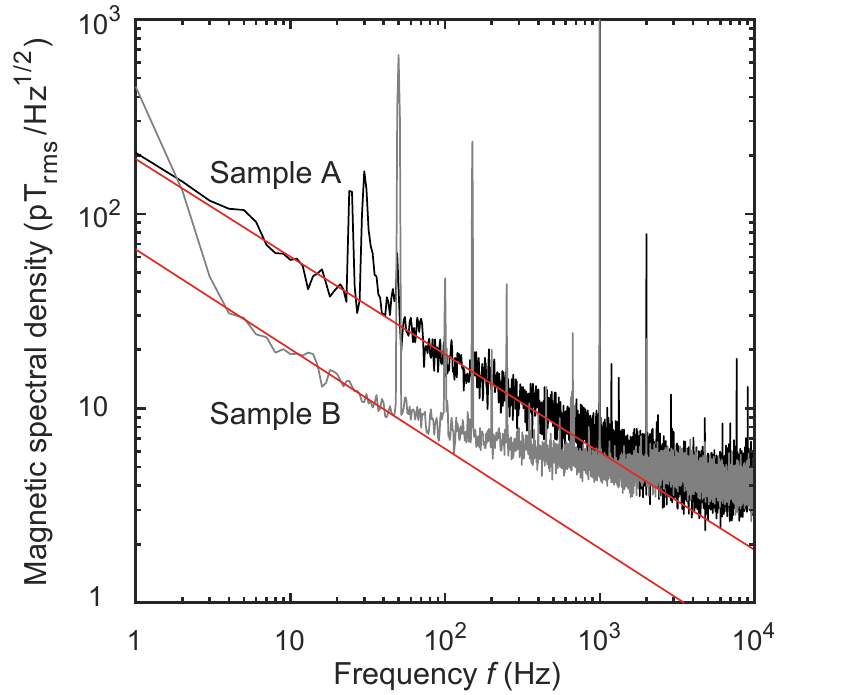}
\caption{\label{fig:Fig3}Measured magnetic noise of Samples A and B at a high readout power of $-19~$dBm, $-46~$dBm, respectively. The fits to the low-frequency noise (lines) are of the form $\propto f^{-0.50}$ with $-0.50$ the best-fit exponent. The longer averaging time of Sample B makes the measurement more susceptible to drifts that are a likely explanation for the noise rise at $f<4$~Hz. The spectral peak at 1~kHz is a deterministic magnetic probe tone, but other peaks are either due to the pick-up of rf interference, or generated by the readout electronics.}
\end{figure}

Finally, we probe the resonances of Samples A and C at a variable temperature. Here, the cooling is mediated by He exchange gas. The extracted $\fr(T)$ and $\Qi(T)$ are the most sensitive to $T$ when the devices are just below $\Tc$ (Fig.~4). We compare the frequency data to an analytical model~\cite{lee_accurate_2005, sato_evaluation_2018}:
\begin{equation}
\frac{\fr(T) - \fr(\Tmin)}{\fr(\Tmin)} \propto \lambda(T=\Tmin) - \lambda_0 \left[ 1- \left( \frac{T}{\Tc} \right)^2 \right]^{-1/2} ,
\end{equation}
where $\Tmin$ is the lowest $T$ in the dataset, and $\lambda_0 = \lambda(T=0)$. For both samples, the best fits of this form have $\Tc =90.5 \pm 0.2$ ~K and $\lambda(T=77~\mathrm{K}) \approx 1.9\lambda_0$ [Fig.~4(a)]. On the other hand, we deduce that $\Qi(T)$ becomes dominantly limited by the intrinsic rf losses when it drops below $10^3$ near $\Tc$ [Fig.~4(b)].
\begin{figure}
\includegraphics{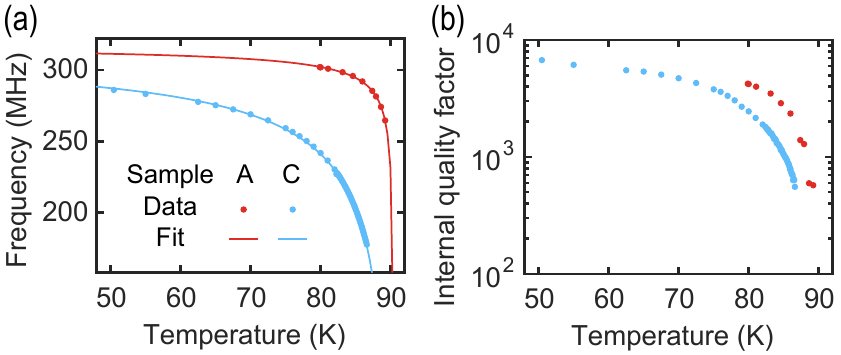}
\caption{\label{fig:Fig4}Measured temperature dependencies of Samples A and C in which the film thicknesses are $d=225~$nm, $50~$nm, respectively. The resonance frequency (a) and the internal quality factor (b) are presented as a function of $T$. See text for the model used for the fits (lines).}
\end{figure}

In conclusion, we have demonstrated high-$\Tc$ kinetic inductance magnetometers with a sensitivity of $4~\mathrm{pT}/\sqrt{\mathrm{Hz}}$ at $10~$kHz and $T=77~$K. They tolerate background fields of $9-28~\mu$T, which is close to the Earth's field. We anticipate that changing the sensor geometry by implementing narrow constrictions~\cite{asfaw_skiffs:_2018} to reduce $\Ic$ should allow for periodical resets, enabling operation at even higher fields. The constrictions should also help to increase the kinetic inductance fraction, a likely route towards a higher device responsivity. Considering the sensitivity, we would find useful a further study of the cause of the low-frequency noise, and methods to minimize it. Future investigations of the YBCO KIM could also evaluate the possibility of beating the 200-mT limit of the perpendicular background field of the NbTiN KIM~\cite{asfaw_skiffs:_2018}.

\begin{acknowledgments}
We thank Maxim Chukharkin for fabricating Sample A, Paula Holmhund for help in sample preparation, and Juho Luomahaara and Heikki Sepp\"a for valuable discussions. V.V., H.S., M.K. and J.H. acknowledge financial support from Academy of Finland under its Center of Excellence Program (project no. 312059), and grants no. 305007 and 310087. S.R., A.K., D.W. and J.S. acknowledge financial support from the Knut and Alice Wallenberg Foundation (KAW2014.0102), the Swedish Research Council (621-2012-3673) and the Swedish Childhood Cancer Foundation (MT2014-0007). We acknowledge support from Swedish national research infrastructure for micro and nano fabrication (Myfab) for device fabrication.
\end{acknowledgments}

\bibliography{HTS_KIM}

\end{document}


\title{Supplement to ``Magnetic field sensing with the kinetic inductance of a high-$\Tc$ superconductor''}
\author{V. Vesterinen}
\email[]{visa.vesterinen@vtt.fi}
\affiliation{VTT Technical Research Centre of Finland Ltd, QTF Center of Excellence, P.O. Box 1000, FI-02044 VTT, Finland}

\author{S. Ruffieux}
\author{A. Kalabukhov}
\affiliation{Department of Microtechnology and Nanoscience—MC2, Chalmers University of Technology, SE-41296 Gothenburg, Sweden}

\author{H. Sipola}
\author{M. Kiviranta}
\affiliation{VTT Technical Research Centre of Finland Ltd, QTF Center of Excellence, P.O. Box 1000, FI-02044 VTT, Finland}

\author{D. Winkler}
\affiliation{Department of Microtechnology and Nanoscience—MC2, Chalmers University of Technology, SE-41296 Gothenburg, Sweden}

\author{J.~F. Schneiderman}
\affiliation{MedTech West and the Institute of Neuroscience and Physiology, University of Gothenburg, SE-40530 Gothenburg, Sweden}

\author{J. Hassel}
\affiliation{VTT Technical Research Centre of Finland Ltd, QTF Center of Excellence, P.O. Box 1000, FI-02044 VTT, Finland}

\date{\today}

\maketitle


\section{Rf coupling configurations}
Our magnetometers are rf resonators that have a considerable size: the edge length of the inductive loop is $W=8$~mm. Hence, there can be unwanted couplings, mediated by rf electromagnetic fields, between the resonator and its surroundings. Therefore it is important to analyze a full rf system consisting of the magnetometer, its connection to an rf feedline, and two concentrically placed low-frequency bias coils. The feedline and the coils are on a sample holder printed circuit board [Fig.~1(b)]. In the worst case, dissipative elements in the vicinity of the magnetometer can limit the maximal attainable quality factor of the resonance. This in turn will affect the device responsivity, i.e., its ability to transduce magnetic field variations into rf domain. 

\begin{figure*}[!htb]
\includegraphics{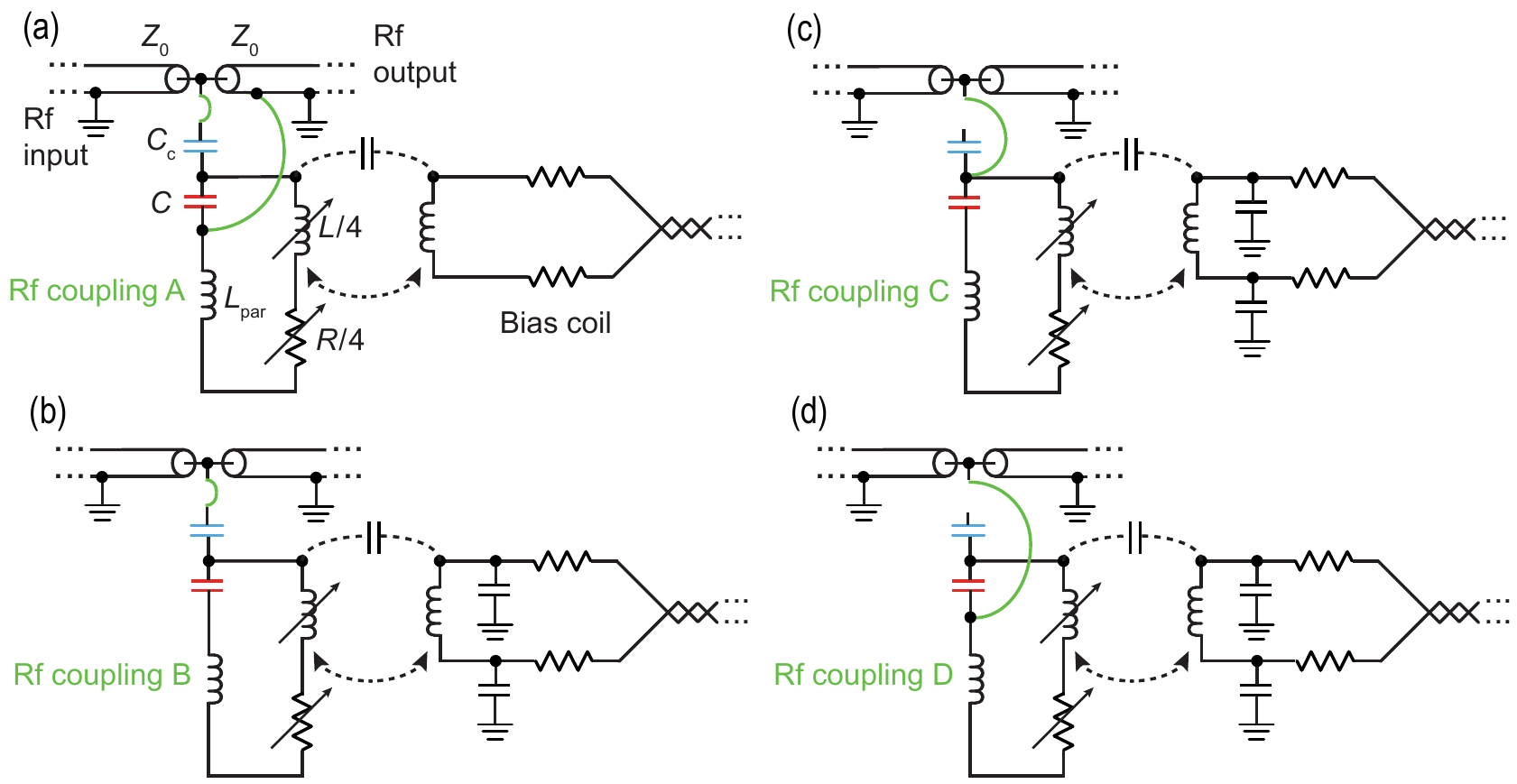}
\caption{\label{fig:FigS4}Rf coupling configurations of the resonator sample. Wirebonds are depicted in green. The rf circuit elements are loop inductance $L$, effective loop resistance $R$, parasitic inductance $\Lpar$ (from a trace connecting the loop to a shunt capacitance $C$), coupling capacitance $\Cc$, and rf characteristic impedance $Z_0$. $R$ represents mainly the intrinsic rf loss of the superconductor.}%
\end{figure*}

To gain insight into the full rf system, we have experimentally tested four rf coupling configurations labeled A-D (Fig.~\ref{fig:FigS4}). They differ from each other in the way the rf return current from the resonator to the sample holder ground is arranged. We change the configuration by rearranging bondwires that connect the magnetometer to the sample holder. Furthermore, some configurations choose to enforce rf grounding of the bias coils, and some choose to control the resonator-feedline coupling strength with a discrete capacitor $\Cc$.

The experimental data presented in Figs.~2-4 have been measured in the following rf coupling configurations: Sample A is in configuration A in Figs.~2-3 with $\Cc=0.9$~pF. For Fig.~4, the coupling of Sample A has been changed to configuration B with $\Cc=0.9$~pF. Sample B is in configuration B in Figs.~2-3 with $\Cc=200$~pF. Sample C is in configuration C in Fig.~4.

Next, we will describe the configurations in detail, discuss the electromagnetic simulation of configurations B-D, and summarize the results of the study. 

\subsection{Configuration A}
In this configuration, bondwire connections are placed between the feedline and the resonator, and between the resonator and the sample holder ground [Fig.~\ref{fig:FigS4}(a)]. The path for the rf return current is thus explicitly provided. Modeling of the coupling strength between the feedline and the resonator is straightforward as it is governed by the coupling capacitor $\Cc$. However, we suspect that stray coupling to the resistive bias coils might lead to an increase of the internal dissipation in the resonator. Series impedance elements are placed to the coils for decoupling purposes.
\subsection{Configuration B}
Like configuration A, but without a bondwire connecting the resonator to the ground [Fig.~\ref{fig:FigS4}(b)]. The rf return current flows through the bias coils. Unlike in configuration A, there is a low-impedance connection between the coils and the ground. The low impedance at radio frequencies is provided by capacitances on the order of nanofarads. These capacitors are too small to influence the flow of the currents generating the magnetic flux bias at low frequencies below 10~kHz. Analyzing the coupling strength between the resonator and the feedline is a complicated task because the coupling is determined by stray electromagnetic fields.
\subsection{Configuration C}
Like configuration B, but with the coupling capacitor bypassed in order to achieve stronger coupling [Fig.~\ref{fig:FigS4}(c)].
\subsection{Configuration D}
Like configuration C, but with the bondwire from the rf feedline connecting to a different location at the resonator [Fig.~\ref{fig:FigS4}(d)]. Again, the rf return current flows through the bias coils.

\subsection{Simulation of configurations B-D}
\begin{figure}[!htb]
\includegraphics[width=242pt]{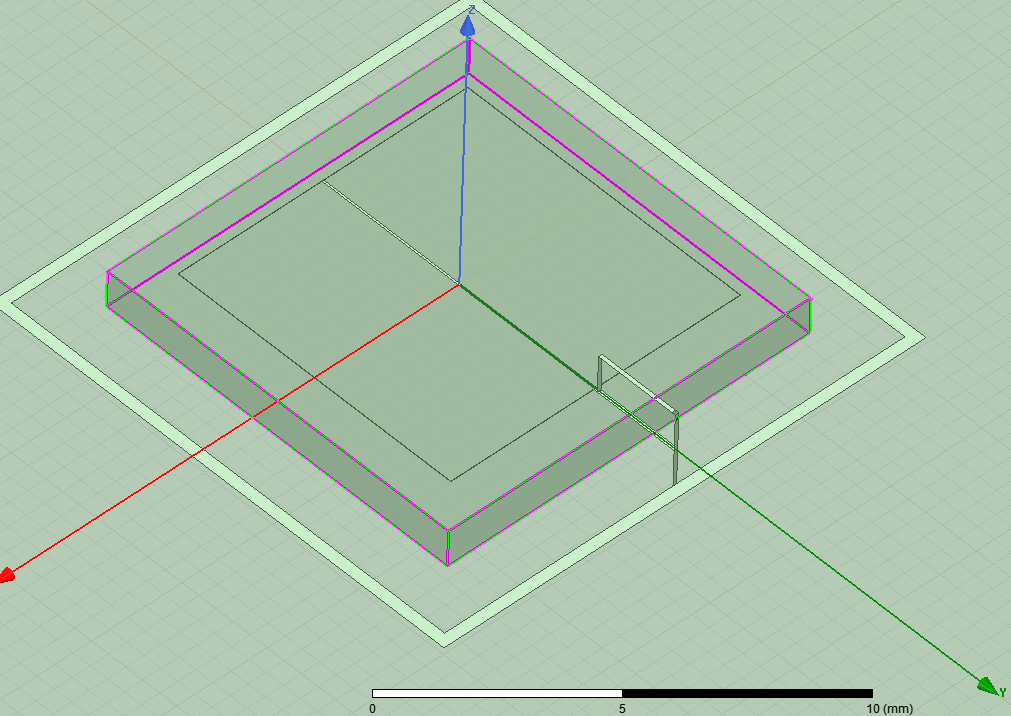}
\caption{\label{fig:FigS5}A screenshot of the simulation model for wirebonding confguration C in Ansys HFSS.}%
\end{figure}

To gain qualitative understanding of the unconventional bondwire configurations B-D, we carry out full-wave electromagnetic simulations in Ansys HFSS. The geometric model is depicted in Fig.~\ref{fig:FigS5}. The simulated magnetometer resembles Sample A. Traces on the magnetometer sample are of actual widths (loop: 10~$\mu$m, parasitic: 30~$\mu$m), and they have a (kinetic) sheet reactance of $6$~m$\Omega$. For simplicity, the simulation neglects optimal rounding~\cite{clem_geometry-dependent_2011,hortensius_critical-current_2012} at the corners and intersections of the loop. The rounding implemented in the actual devices should prevent critical current crowding effects. The interdigital shunt capacitor $C$ of the resonator is approximated as a lumped element. We add some internal dissipation to the resonator with a 100-k$\Omega$ lumped resistor placed in parallel to $C$. The size and permittivity of the magnetometer substrate are taken into account, but the loss tangent of the substrate is set to zero.

The sample holder is a 0.8-mm-thick FR-4 board ($\epsr=4.4$, loss tangent 0.02) floating in an open space of vacuum. For simplicity, we simulate only one bias coil with a single turn. It has a trace width of 0.3~mm, and an edge length of 13~mm that corresponds to the average size of the actual, three-turn dc bias coil. The losses of the copper trace are modeled with a sheet resistance of $0.14$~m$\Omega$. An rf excitation port interrupts a bondwire that connects the coil to the magnetometer. The bondwire is approximated as sections of an infinitesimally thin, 100~$\mu$m wide sheet.  Furthermore, the coupling capacitor $\Cc$ of configuration B is embedded into the bondwire as a lumped element. 

\begin{figure}[!htb]
\includegraphics{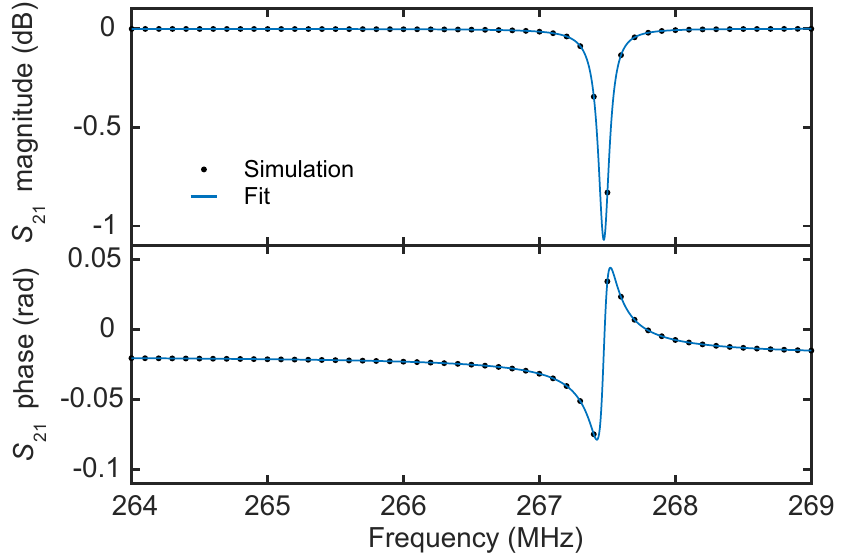}
\caption{\label{fig:FigS8}Simulated resonance lineshape of Sample A in rf coupling configuration B.}
\end{figure}

We fine-tune $C$ to produce a resonance frequency of approximately 270~MHz in the simulation, which is close to the value measured for Sample A. The output of the simulation is a dataset of the impedance $\Zres$ (seen from the excitation port) \emph{vs.} radio frequency, and we convert it into an equivalent transmission S-parameter
\begin{equation}
S_{21} = \frac{2\Zres}{2\Zres+Z_0},
\end{equation}
where $Z_0=50$~$\Omega$ is the characteristic rf impedance. From the simulated transmission parameter (Fig.~\ref{fig:FigS8}) we extract the resonance frequency and the quality factors using the same fitting routine that is used for the experimental data.

\subsection{Reference samples from low-$\Tc$ superconductors}
Resonator samples with a minimal amount of internal dissipation are necessary for evaluating the performance limits of the sample holder and the different rf coupling configurations. To this end, we have fabricated additional magnetometers from low-$\Tc$ superconducting films sputter-deposited onto high-resistivity silicon substrates. We study both niobium~\cite{gronberg_side-wall_2017} (thickness $d=200$~nm) with a low kinetic inductance fraction $\alphak$, and niobium nitride~\cite{luomahaara_kinetic_2014} ($d=165$~nm) with a high $\alphak$. The layouts of the low-$\Tc$ samples differ from those of the high-$\Tc$ samples in certain geometrical aspects: they have a narrower trace width of $w=5$~$\mu$m within the loop, and modified dimensions of the interdigital capacitors. However, the resonance frequency of the NbN magnetometer is similar to that of the high-$\Tc$ Samples A and B.

\subsection{A comparison of the configurations}
Table~\ref{table:lineshapes} summarizes the effects of the variable rf coupling configuration. We have measured the high-$\Tc$ Sample A in all four configurations at a temperature of $T=77-80$~K, and simulated it in configurations B-D. The low-$\Tc$ samples have been measured at $T=4.2$~K in configurations B and D.

With Sample A we are able reach higher values of the internal quality factor $\Qi$ by moving from configuration A into configurations B-D. Hence, it seems beneficial to guide the rf return current through the bias coils. We measure that the $\Qi$ of Sample A can increase by a factor as large as two, despite studying the configurations B-D at a higher temperature where we are more susceptible to the superconductor internal loss. When comparing the unconventional configurations B-D, we observe with all the samples that D provides the strongest coupling to the rf feedline. However, the configurations B-C are more favorable in terms of achieving a high $\Qi$.

We note that making the coupling between the resonator and the feedline weaker has the effect of pushing the resonance frequency $\fr$ to a higher value. Based on this relation we suspect that all configurations produce a net capacitive coupling between the magnetometer and the feedline. The simulated coupling strength, expressed in terms of the external quality factor $\Qe$, is in good qualitative agreement with the measurements. The simulation overestimates $\Qe$ by a factor of less than two, which we attribute to the simplified simulation geometry. The simulation omits the second bias coil as well as the presence of a ground plane on the sample holder at a lateral distance of about 5~mm from the superconducting loop [Fig.~1(b)].

\begin{figure}[!htb]
\includegraphics{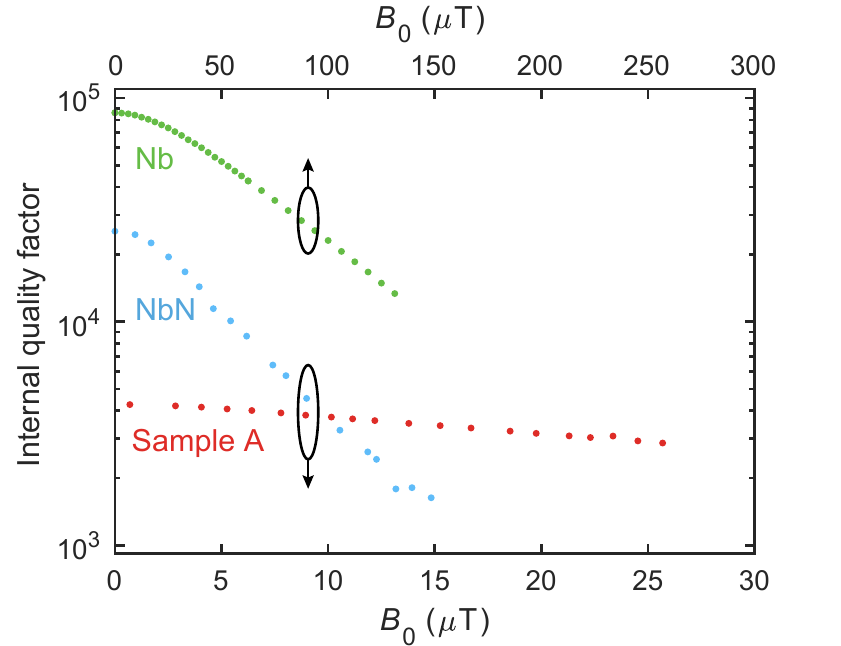}
\caption{\label{fig:FigS6}Measured resonator internal quality factor as a function of the applied magnetic flux density $B_0$, for three types of samples in rf coupling configuration B. The $B_0$-induced frequency shifts translate into maximal inductance variations of $0.14\%$, $0.03\%$ and $2\%$ in YBCO (Sample A, red), Nb (green) and NbN (blue), respectively.}
\end{figure}

As a conclusion, we think that configuration B is well-suited for sensitive magnetometry. This configuration combines the highest $\Qi$ with an adjustable $\Qe$. We note that variations in the coupling strength can be realized by placing some or all of the coupling capacitance onto the sample holder in the form of rf-compatible, surface-mount components.  We present in Fig.~\ref{fig:FigS6} for all three samples the measurement of $\Qi$ as a function of the magnetic flux density $B_0$ generated by a current applied to the dc bias coil. Here, the rf coupling configuration is B, and further details are listed in Table~\ref{table:lineshapes}.

\begin{table*}[!htb]
\caption{Summary of measured and simulated resonator lineshapes at zero applied magnetic flux.}
\begin{tabular}{l|c|c|r|}
Sample                 & Rf coupling configuration & Environment & Lineshape\\
\hline
(A) YBCO $d=225$~nm        & A with $\Cc=0.9$~pF & $T=77$~K & $\fr=285.8$~MHz, $\Qi=2100$, $\Qe=350$\\
(A) YBCO $d=225$~nm        & B with $\Cc=0.9$~pF& $T=80$~K & $\fr=302.1$~MHz, $\Qi=4300$, $\Qe=15700$\\
&&& Simulated $\Qe=22000$\\
(A) YBCO $d=225$~nm      	& C & $T=80$~K & $\fr=301.3$~MHz, $\Qi=4200$, $\Qe=3900$\\
&&& Simulated $\Qe=6200$\\
(A) YBCO $d=225$~nm	& D & $T=80$~K  & $\fr=298.7$~MHz, $\Qi=3050$, $\Qe=930$\\
&&& Simulated $\Qe=1760$\\
\hline
Nb $d=200$~nm		& B with $\Cc=0.4$~pF& $T=4.2$~K & $\fr=358.9$~MHz, $\Qi=86000$, $\Qe=28000$\\
Nb $d=200$~nm		& D & $T=4.2$~K & $\fr=346.2$~MHz, $\Qi=16000$, $\Qe=700$ \\
\hline
NbN $d=165$~nm		& B with $\Cc=0.4$~pF& $T=4.2$~K & $\fr=275.2$~MHz, $\Qi=25000$, $\Qe=34000$\\
NbN $d=165$~nm		& D & $T=4.2$~K & $\fr=271.0$~MHz, $\Qi=13000$, $\Qe=1400$ \\
\hline
\end{tabular}
\label{table:lineshapes}
\end{table*}

\section{Wiring diagrams}
\begin{figure*}[!htb]
\includegraphics{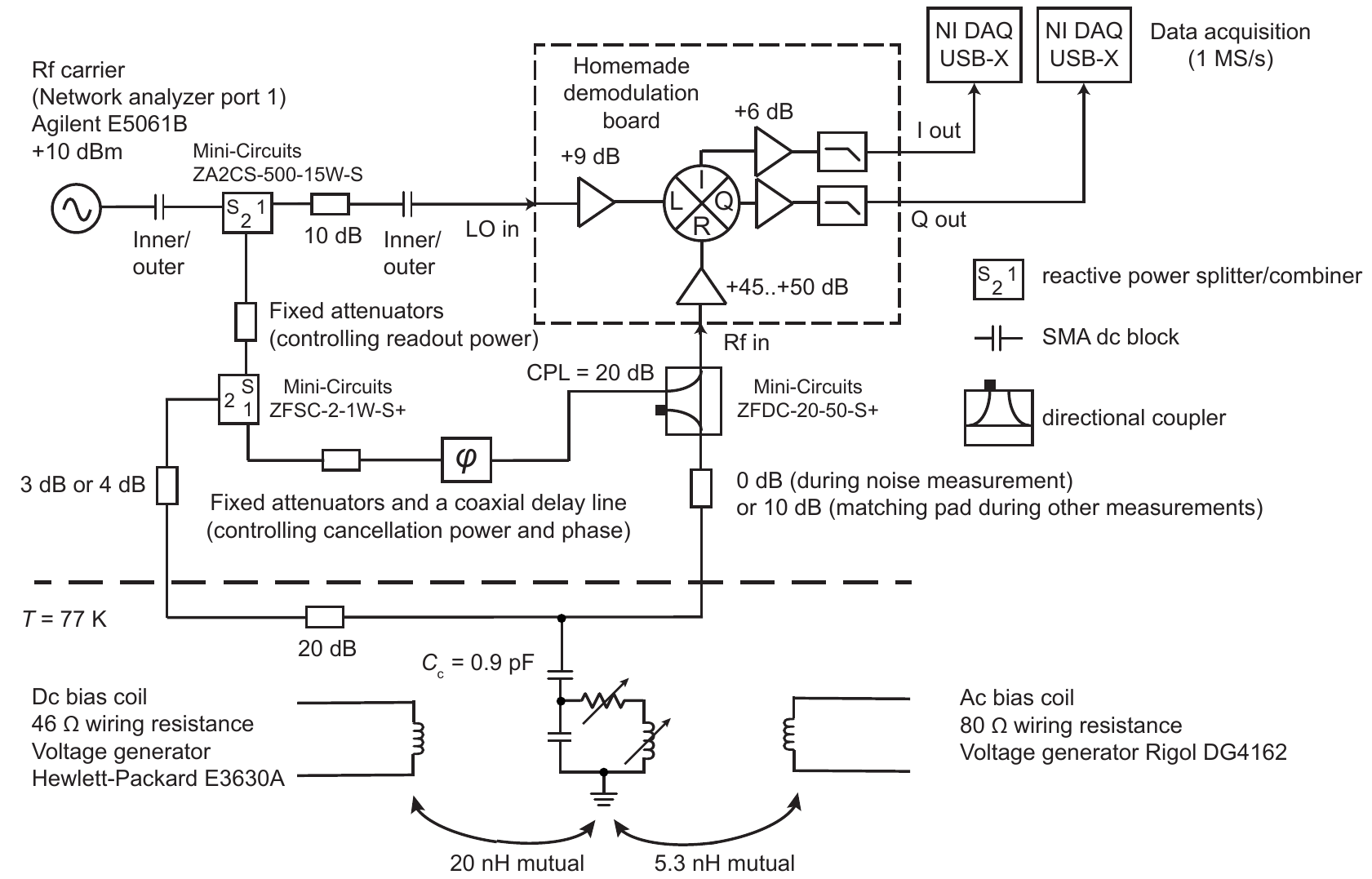}
\caption{\label{fig:FigS1}Detailed wiring diagram of Sample A for measurements presented in Figs.~2-3 and Fig.~\ref{fig:FigS3}. We use the rf coupling configuration configuration A. Readout rf signals are amplified and demodulated on a homemade four-layer printed circuit board. All components on the board are of surface mount type: the rf amplifiers are from Mini-Circuits, and the IQ demodulator and the intermediate frequency amplifiers from Analog Devices. The low-pass filters are first-order $RC$ circuits with a corner frequency of 16~kHz.}%
\end{figure*}

\begin{figure*}[!htb]
\includegraphics{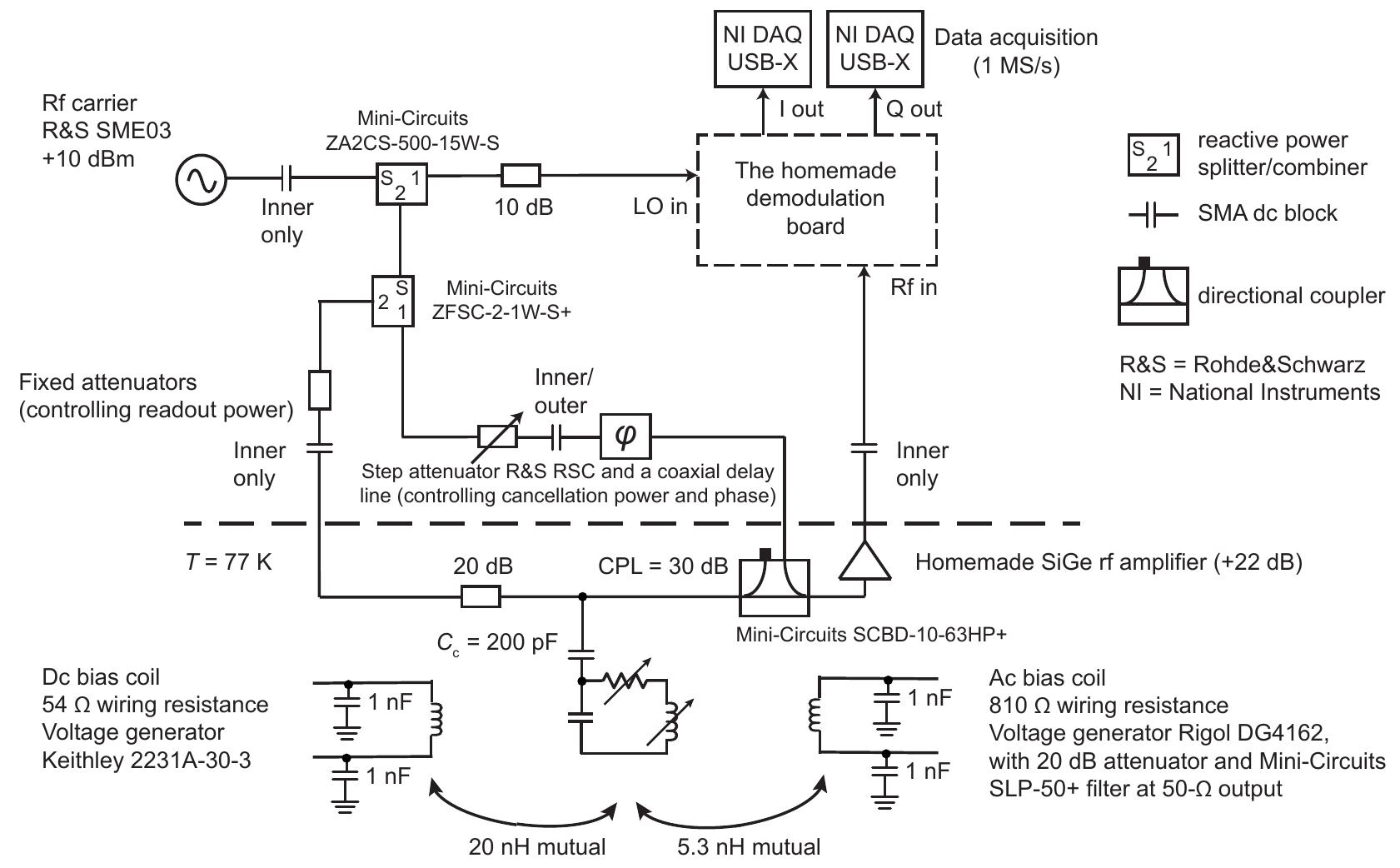}
\caption{\label{fig:FigS2}Detailed wiring diagram of Sample B for measurements presented in Figs.~2-3 and Fig.~\ref{fig:FigS3}. We use the rf coupling configuration B.}%
\end{figure*}

The detailed wiring diagrams for the time-domain measurements of Sample A and B (Figs.~2-3 and Fig.~\ref{fig:FigS3}) are in Fig.~\ref{fig:FigS1} and Fig.~\ref{fig:FigS2}, respectively. We carried out frequency-domain characterization of the magnetometer resonances (Fig.~4, Fig.~\ref{fig:FigS6}, Table~\ref{table:lineshapes}) with a a vector network analyzer.

\subsection{Cryogenic SiGe rf preamplifier}
\begin{figure}[!htb]
\includegraphics{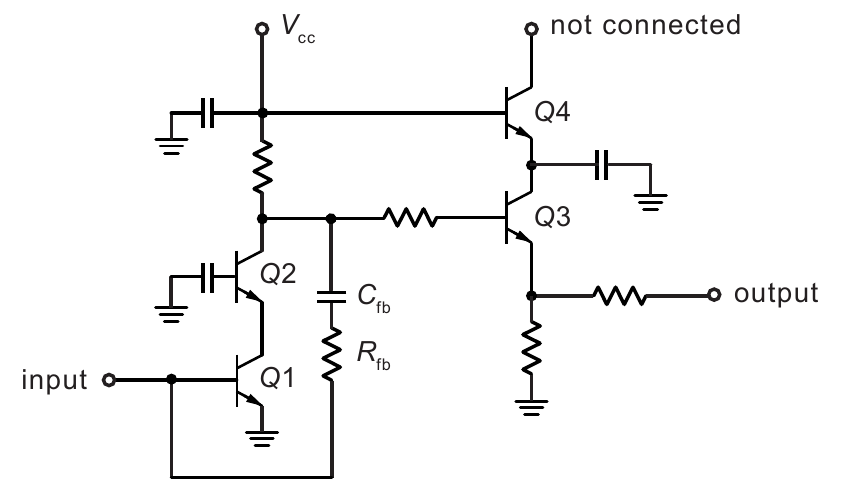}
\caption{\label{fig:FigS7}Simplified rf schematic of the SiGe transistor amplifier. Elements omitted from the diagram include dc blocking capacitors at input and output, as well as resistor networks for setting the dc bias points of the transistors.}
\end{figure}

Our custom-designed cryogenic amplifier has four SiGe heterojunction bipolar transistors~\cite{kiviranta_use_2006}, labeled $Q1-Q4$ in Fig.~\ref{fig:FigS7}. It is designed for the radio frequency range of $10-1000$~MHz. It is built of low-cost, commercially available surface-mount components on a two-sided FR-4 printed circuit board with a thickness of 0.8~mm. The amplifier occupies an area of $20$~mm~$\times$~$15$~mm on the board, plus the footprint of input/output SMP connectors.

We describe the roles of the transistors $Q1-Q4$, all of which are the model BFP 640F H6327 from Infineon. The cryo-compatibility of this model has previously been reported in Ref.~\onlinecite{ivanov_microwave_2016}. In our design, the stage providing gain is the cascode formed by a common-emitter $Q1$ and a common-base $Q2$. The amount of gain is controlled with the impedance seen by the $Q2$ collector. We add $Q3$ as a common-collector buffer stage. Finally, the role of $Q4$ is to provide a dc voltage drop: the cryogenic base-emitter voltage of SiGe is about 1~V. Broadband input (output) match to $Z_0=50$~$\Omega$ is achieved with feedback consisting of a resistor $R_\mathrm{fb}$ and a capacitor $C_\mathrm{fb}$ (the choice of resistors at the $Q3$ emitter).

At 77~K temperature, we use a supply voltage of $V_\mathrm{cc}\simeq 3.5$~V and the amplifier dissipates about $27$~mW.

\section{Extended data}
\begin{figure}[!htb]
\includegraphics[width=242pt]{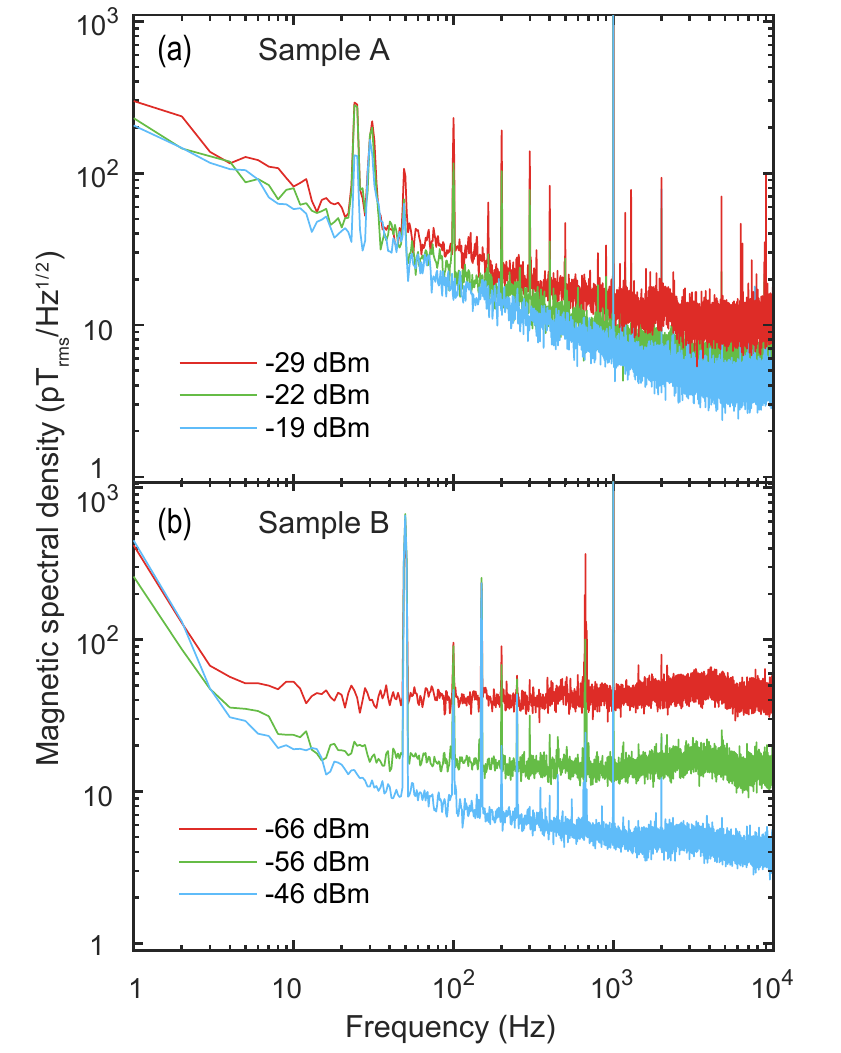}
\caption{\label{fig:FigS3}Measured magnetic noise of Samples A and B. The rf readout power levels arriving at the samples are indicated. The data at the highest power levels (blue) are the same as those presented in Fig.~3.}%
\end{figure}

We present in Fig.~\ref{fig:FigS3} the magnetic field noise of Samples A and B as a function of the rf readout power level. 

\bibliographystyle{prl}
\bibliography{HTS_KIM}